\documentclass[aps,prc,reprint,superscriptaddress,showpacs]{revtex4-1}

\usepackage{color}   
\usepackage{graphicx}
\usepackage{dcolumn} 
\usepackage{bm}      
\usepackage{slashed}
\usepackage{amssymb}
\usepackage{amsmath}

\begin{document}


\title{Toroidal high-spin isomers in the nucleus $^{304}{120}$}



\author{A. Staszczak}
\affiliation{Institute of Physics, Maria Curie-Sk{\l}odowska University,
pl.~M.~Curie-Sk{\l}odowskiej 1, 20-031 Lublin, Poland}

\author{Cheuk-Yin~Wong}
\affiliation{Physics Division, Oak Ridge National Laboratory, Oak~Ridge, TN 37830, USA}

\author{A. Kosior}
\affiliation{Institute of Physics, Maria Curie-Sk{\l}odowska University,
pl.~M.~Curie-Sk{\l}odowskiej 1, 20-031 Lublin, Poland}


\date{\today}

\begin{abstract}
\begin{description}
\item[Background]
Strongly deformed oblate superheavy nuclei form an intriguing region where
the toroidal nuclear structures may bifurcate from the oblate spheroidal shape.
The bifurcation may be facilitated when the nucleus is endowed with a large
angular moment about the symmetry axis with $I=I_{z}$.
The toroidal high-$K$ isomeric states at their local energy minima can be
theoretically predicted using the cranked self-consistent Skyrme-Hartree-Fock method.
\item[Purpose]
We use the cranked Skyrme-Hartree-Fock method to predict the properties of the
toroidal high-spin isomers in the superheavy nucleus $^{304}{120}_{184}$.
\item[Method]
Our method consists of three steps: first, we use the deformation-constrained
Skyrme-Hartree-Fock-Bogoliubov approach to search for the nuclear density
distributions with toroidal shapes. Next, using these toroidal distributions
as starting configurations we apply an additional cranking constraint of
a large angular momentum $I=I_{z}$ about the symmetry $z$-axis and search for
the energy minima of the system as a function of the deformation.
In the last step, if a local energy minimum  with $I=I_{z}$ is found, we perform
at this point the cranked symmetry- and deformation-unconstrained
Skyrme-Hartree-Fock calculations to locate a stable toroidal high-spin isomeric
state in free convergence.
\item[Results]
We have theoretically located two toroidal high-spin isomeric states of $^{304}{120}_{184}$
with an angular momentum $I$=$I_z$=81$\hbar$ (proton 2p-2h, neutron 4p-4h excitation)
and $I$=$I_z$=208$\hbar$ (proton 5p-5h, neutron 8p-8h) at the quadrupole moment deformations
$Q_{20}=-297.7$~b and $Q_{20}=-300.8$~b with energies 79.2 MeV and 101.6 MeV above the
spherical ground state, respectively.
The nuclear density distributions of the toroidal high-spin isomers
$^{304}{120}_{184}(I_z$=81$\hbar$ and 208$\hbar$) have the maximum density close
to the nuclear matter density, 0.16 fm$^{-3}$, and a torus major to minor radius
aspect ratio $R/d=3.25$.
\item[Conclusions]
We demonstrate that aligned angular momenta of $I_z$=81$\hbar$ and 208$\hbar$
arising from multi-particle-multi-hole excitations in the toroidal system of
$^{304}{120}_{184}$ can lead to high-spin isomeric states, even though the toroidal
shape of $^{304}120_{184}$ without spin is unstable.
Toroidal energy minima without spin may be possible for superheavy nuclei with
higher atomic numbers, $Z\gtrsim$122, as reported previously \cite{Sta08}.
\end{description}
\end{abstract}



\maketitle

%
\section{Introduction}

\begin{figure}[htbp]
\begin{center}
\includegraphics[width=\columnwidth]{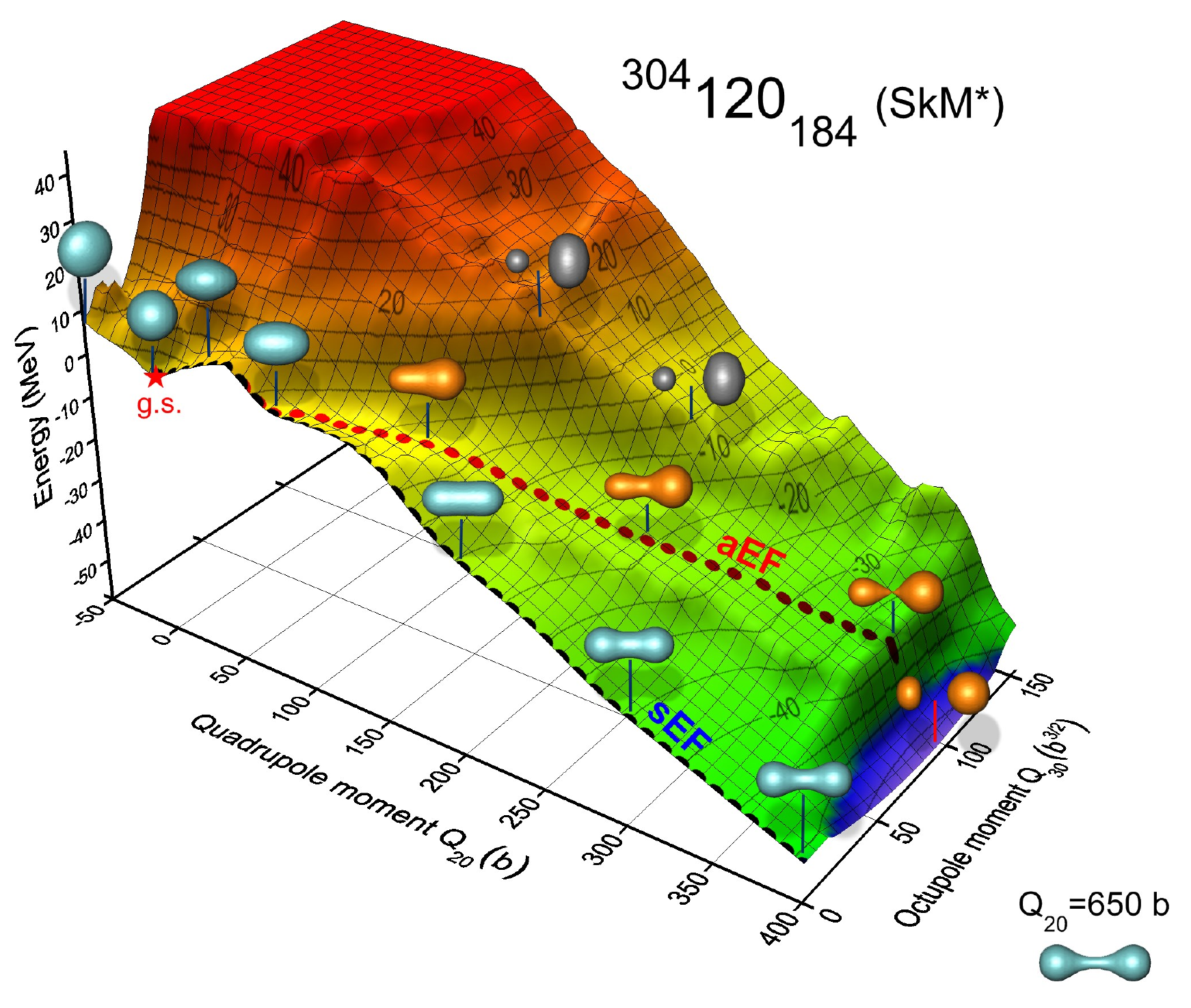}
  \caption{\label{Fig1} (Color online) Total HFB energy surface of
  $^{304}120_{184}$  as a function of the quadrupole $Q_{20}$ and
  octupole $Q_{30}$ moments. The HFB energy is normalized to the
  ground state energy.
  The dashed lines show the symmetric (sEF) and asymmetric elongated
  fission (aEF) paths along different valleys.}
\end{center}
\end{figure}

The landscape of the total energy surface of a nucleus in the deformation
degrees of freedom is central to our understanding of the equilibrium
shapes and the evolutionary paths in nuclear dynamics. In Fig.~\ref{Fig1}
one can see the total energy surface for the superheavy nucleus $^{304}120_{184}$
as a function of the quadrupole and octupole degrees of freedom calculated
in the constrained Hartree-Fock-Bogoliubov (HFB) approach with the Skyrme
energy density functional.
In addition to the spherical ground state minimum, the landscape contains
the symmetric-elongated-fission (sEF) and asymmetric-elongated-fission
(aEF) paths leading to fission. These features have important experimental
implications in the multimodal fission decay properties of heavy and superheavy
nuclei (\textit{cf.} Refs.~\cite{Sta09,War12,Sta13}).

The potential energy surface in Fig.~\ref{Fig1} pertains to reflection-symmetric
and reflection-asymmetric prolate shapes. How does the energy surface look like
in the oblate deformation region? What kinds of the nuclear (equilibrium) shapes
may there be in this oblate deformation region?

\begin{figure}[htbp]
\begin{center}
\includegraphics[width=\columnwidth]{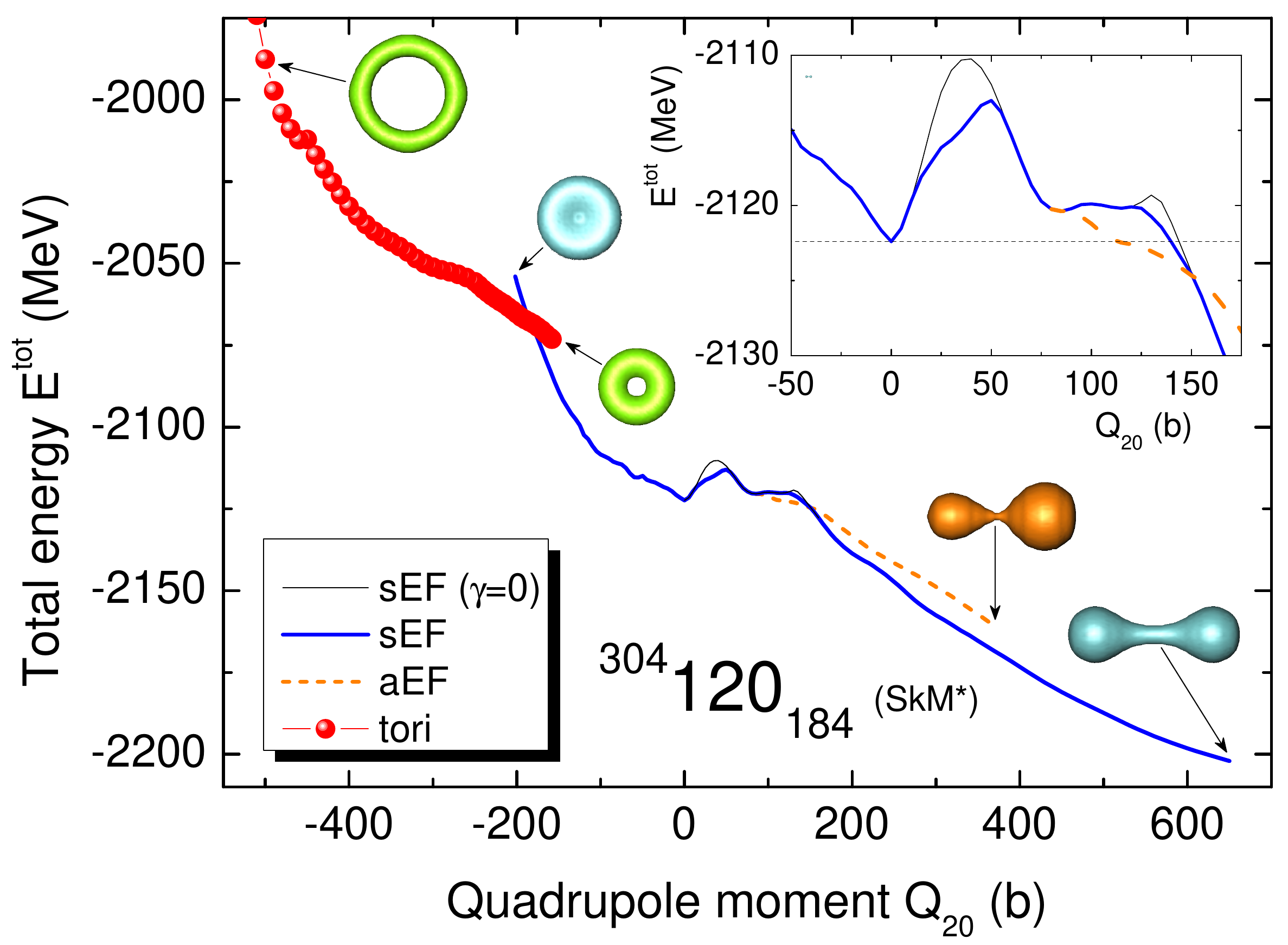}
  \caption{\label{Fig2} (Color online) Total HFB energy curve of $^{304}120_{184}$
  as a function of the quadrupole moment.
  The thick solid (blue color) and gray dashed (orange) lines show the symmetric (sEF)
  and asymmetric (aEF) elongated fission pathways along different valleys, respectively.
  The effect of triaxiality on the inner and outer barrier is shown in the inset,
  where the axially symmetric sEF ($\gamma = 0^{\circ}$) fission pathway is marked
  by solid thin (black) line. The nuclear matter density distributions with
  toroidal shapes appear at the region of large oblate deformation
  $Q_{20} \leqslant - 158$ b dark gray (red) solid circles.}
\end{center}
\end{figure}

To gain the proper perspective, it is informative to discuss some general features
of our results in the prolate and oblate regions and then examine in details in this
paper how oblate region results are obtained.
The total HFB energy of $^{304}120_{184}$ as a function of the quadrupole moment
$Q_{20}$ is shown in Fig.~\ref{Fig2}. On the prolate deformation side,
the pre-scission density configurations for the sEF and aEF paths are shown at the
ends of both paths (at $Q_{20}\thickapprox 360$~b for sEF and $Q_{20}\thickapprox 650$~b
for aEF). The effects of triaxiality on the change of the inner and outer axial-symmetric
barriers are shown in the insert of Fig.~\ref{Fig2}.
On the oblate deformation side with a negative $Q_{20}$, one starts from the energy
minimum for a spherical ground state to go to the higher energies for oblate spheroids.
As the oblate $Q_{20}$ magnitude increases, the oblate spheroidal density changes into
a biconcave disc with flattened center density.
At $Q_{20}\thickapprox - 200$ b, the biconcave disc energy surface reaches an energy
about 72 MeV above the spherical ground state.
Upon a further increase in the oblate deformation a sudden shape transition from
a biconcave disc to a torus takes place with a reduction of the total energy of
the nucleus by 10.8 MeV.

The geometry of the toroidal nuclear densities can be characterized by
the aspect ratio $R/d$, where  $R$ is the major radius, the distance from
the center of the torus hole to the center of the torus tube, and $d$ is
the minor radius, the radius of the tube. As is shown in Fig.~\ref{Fig2}
for $Q_{20}\le$ -158 b, the aspect ratio $R/d$ of the toroidal solution of
the Skyrme-HFB model increases as the oblate $Q_{20}$ magnitude increases.

With regard to  the emergence of toroidal nuclear matter densities, Wheeler
suggested long ago that under appropriate conditions the nuclear fluid may
assume a toroidal shape \cite{Gam61}. Conditions that are favorable for the
formation of nuclei with a toroidal shape are the cases of excess charge, excess angular momentum,
and nuclear shell effects \cite{Won73,Won78}. In the semi-classical liquid-drop
model, nuclei with a toroidal shape begin to develop as the fissility parameter $x$ exceeds 0.964.
However, the toroidal nucleus is plagued with various instabilities~\cite{Won73},
and the search for toroidal nuclei continues \cite{Sta08}. When a toroidal nuclear system
is endowed with an angular momentum along the symmetry axis, $I$=$I_z$,
the variation of the rotational energy of the spinning nucleus can counterbalance
the variation of the toroidal bulk energy to lead to toroidal isomeric
states at their local energy minima, when the angular momentum $I$=$I_z$
is beyond a threshold value~\cite{Won78}. A rotating liquid-drop toroidal nucleus can
also be stable against sausage instabilities (know also as Plateau-Rayleigh
instabilities, in which the torus breaks into smaller fragments~\cite{Egg97,Pai09}),
when the same mass flow is maintained across the toroidal meridian to lead
to high-spin isomers within an angular momentum window~\cite{Won78}.

The rotating liquid-drop model is useful as an intuitive, qualitative guide
to point out the essential balance of forces leading to possible toroidal
figures of equilibrium. The quantitative assessment of toroidal high-spin
isomer (THSI) relies on microscopic descriptions that include both the bulk
properties of the nucleus and the single-particle shell effects in self-consistent
mean-field theories, such as the Skyrme-Hartree-Fock (Skyrme-HF) approach.
Self-consistent mean-field theories are needed because non-collective rotation
with an angular momentum about the symmetry axis is permissible quantum mechanically
for an axially symmetric toroid only by making single-particle particle-hole
excitations to align the angular momenta of the constituents along the symmetry axis
\cite{And76,Boh81,Ring80,Voi83,Nils95,Afa99}. As a consequence, only a certain
discrete, quantized set of total angular momentum $I$=$I_z$ states is allowed.
We shall adopt the simplified notation that all spins and angular momenta
are implicitly in units of $\hbar$ except otherwise explicitly indicated to
resolve ambiguities.

It was recently found that the THSI with $I$=60 may
be in the local energy minimum in the excited states of $^{40}$Ca by
using a cranked Skyrme-HF method starting from the initial ring configuration
of 10 alpha particles \cite{Ich12,Ich14}.
Using a cranked Skyrme-HF approach, it was found  that toroidal high-spin
isomeric states have actually a rather general occurrence for an extended
region of even-even light nuclei with 28$\leqslant$$A$$\leqslant$52 \cite{Sta14}.
With different rings of alpha particles as initial states, it was also
subsequently confirmed that there are THSIs solutions
in the extended region of 36$\leqslant$$A$$\leqslant$52 \cite{Ich14a}.
The particle-hole nature of the light high-spin toroidal isomers has been
examined in Ref.~\cite{Sta15a}, the toroidal high-spin isomers with $N\ne Z$
have been located \cite{Sta15b}, and the THSIs in $^{56}$Ni have been described
in Ref.~\cite{Sta16}. For the nucleus $^{24}$Mg, a toroidal diabetic excited
state without spin has also been found \cite{Zha10}.

In addition to the high-spin toroidal isomers in the light mass region,
the superheavy nuclei with large atomic numbers provides another favorable
region for toroidal nuclei formation, because the large Coulomb repulsion
tends to push the nuclear matter outward to make it energetically advantageous
to assume a toroidal shape.
A previous work in the superheavy region in the self-consistent constraint
Skyrme Hartree-Fock+BCS (Skyrme-HF+BCS) framework indicates that toroidal energy
minima are present at various energies as the atomic number increases beyond
$Z\agt 122$ \cite{Sta08}.
For example, the superheavy nuclei $^{316}122_{194}$, $^{340}130_{210}$, and
$^{352}134_{218}$ have toroidal local potential energy minima lying at about
50, 25, and 12 MeV above their corresponding deformed oblate ground state
energy minimum, respectively. The superheavy nucleus $^{364}138_{226}$ has
a toroidal local potential energy minimum that lies even below the oblate
spheroidal energy minimum.

Our purpose in the present manuscript is to explore the closed-shell superheavy
nucleus $^{304}$120$_{184}$ which is localized close to the center of the
``island of stability''(\textit{cf.} Refs.~\cite{Sta13,Bar15}).
Toroidal system of $^{304}$120$_{184}$ without a spin may not be stable \cite{Kosi17}.
It is of interest to find out whether the superheavy nucleus $^{304}$120$_{184}$
with a toroidal density may become stabilized by the addition of a large nuclear spin.

This manuscript is organized as follows. In Secs.~II~A-C, we describe the
theoretical model. In Sec.~III~A, we examine properties of the toroidal system of
$^{304}120_{184}$ without spin and study the single-particle states in the
constrained Skyrme-HFB calculations as a function of the quadrupole moment.
In Sec.~III~B, we present results of the cranked Skyrme-HF calculations
for $^{304}120_{184}$ with a toroidal density and a spin. The properties of
$^{304}120_{184}$ toroidal high-spin isomers are presented in Sec.~III~C.
Finally, we summarize our studies in Sec.~VI.

\section{Description of the model}
\subsection{The Skyrme energy density functional}
In the local density approximation the Skyrme energy density functional (EDF),
up to second-order in derivatives of the density (\emph{i.e.}, the most general
quadratic EDF), can be expressed in terms of seven proton and neutron
local densities:
the particle (scalar) density $\rho_{q}(\boldsymbol{r})$,
kinetic energy (scalar) density $\tau_{q}(\boldsymbol{r})$,
spin-current (pseudotensor) density $\mathbb{J}_{q}(\boldsymbol{r})$,
current (vector) density $\boldsymbol{j}_{q}(\boldsymbol{r})$,
spin (pseudovector) density $\boldsymbol{s}_{q}(\boldsymbol{r})$,
spin-kinetic (pseudovector) density $\boldsymbol{T}_{q}(\boldsymbol{r})$, and
tensor-kinetic (pseudovector) density $\boldsymbol{F}_{q}(\boldsymbol{r})$,
where $q=\{p,n\}$, see Refs.~\citep{Perl04,Lesi07,Bend09,Hell12}.

The above local densities are all real, and $\rho_{q}(\boldsymbol{r})$,
$\tau_{q}(\boldsymbol{r})$, and $\mathbb{J}_{q}(\boldsymbol{r})$ are time-even,
whereas
$\boldsymbol{j}_{q}(\boldsymbol{r})$, $\boldsymbol{s}_{q}(\boldsymbol{r})$,
$\boldsymbol{T}_{q}(\boldsymbol{r})$, and $\boldsymbol{F}_{q}(\boldsymbol{r})$
are time-odd.
The spin-current pseudotensor density $\mathbb{J}_{q}(\boldsymbol{r})$ can
be decomposed into trace, antisymmetric and symmetric parts, giving the pseudoscalar
$\mathcal{J}_{q}(\boldsymbol{r})$, vector $\boldsymbol{J}_{q}(\boldsymbol{r}),$
and (traceless) pseudotensor $\mathfrak{J}_{q}(\boldsymbol{r})$ densities, respectively.

The time reversal and spatial symmetries impose restrictions on the local
densities \citep{Vau72,Roho10}.
In spherical nuclei (the rotational and mirror symmetry, $\mathrm{O(3)}$)
the pseudoscalar $\mathcal{J}_{q}(\boldsymbol{r})$, all the pseudovector
($\boldsymbol{s}_{q}(\boldsymbol{r})$, $ \boldsymbol{T}_{q}(\boldsymbol{r})$,
$\boldsymbol{F}_{q}(\boldsymbol{r})$) and the pseudotensor
$\mathfrak{J}_{q}(\boldsymbol{r})$ local densities vanish.
In the case of axial- and reflection-symmetry only the pseudoscalar
component $\mathcal{J}_{q}(\boldsymbol{r})$ vanishes.
For the description of static properties in even-even nuclei, all the time-odd
densities must vanish to preserve the time-reversal invariance of the density matrix
in the particle-hole channel.

The standard proton-neutron separable Skyrme EDF can be divided into two parts,
built of the seven isoscalar ($t$=0) and seven isovector $T_z$=0 component ($t$=1)
single-particle densities \cite{Vau72}
\begin{equation}
E_{Sk}= \sum_{t=0,1}\int \mathrm{d}^{3} \boldsymbol{r}
\left(\mathcal{H}_{t}^{even}(\boldsymbol{r})+
\mathcal{H}_{t}^{odd}(\boldsymbol{r})\right),
\label{eq:3}
\end{equation}
where the isoscalar densities are the total ($n+p$) densities, while the
isovector densities are the differences of the neutron and proton ($n-p$) densities.
The energy densities $\mathcal{H}_{t}^{even}(\boldsymbol{r})$ and
$\mathcal{H}_{t}^{odd}(\boldsymbol{r})$ are the real, time-even, scalar, and isoscalar
functions of local densities and their derivatives.
The time-even part of Skyrme EDF
\begin{eqnarray}
\mathcal{H}_{t}^{even}(\boldsymbol{r})&=& C^{\rho}_{t}[\rho_{0}]\rho^{2}_{t}
+ C^{\Delta\rho}_{t}\rho_{t}\Delta\rho_{t}+ C^{\tau}_{t}\rho_{t}\tau_{t}\nonumber \\
&&+ C^{J0}_{t}\mathcal{J}^{2}_{t}+ C^{J1}_{t}\boldsymbol{J}^{2}_{t}+
C^{J2}_{t}\mathfrak{J}^{2}_{t}\nonumber \\
&&+ C^{\nabla J}_{t}\rho_{t} \boldsymbol{\nabla}\cdot\boldsymbol{J}_{t},
\label{eq:4}
\end{eqnarray}
is expressed as a bilinear form of the time-even densities and their derivatives.
The time-odd Skyrme EDF
\begin{eqnarray}
\mathcal{H}_{t}^{odd}(\boldsymbol{r})&=& C^{s}_{t}[\rho_{0}]\boldsymbol{s}^{2}_{t}
+ C^{\Delta s}_{t}\boldsymbol{s}_{t}\cdot\Delta\boldsymbol{s}_{t}
+ C^{T}_{t}\boldsymbol{s}_{t}\cdot\boldsymbol{T}_{t}+ C^{j}_{t}\boldsymbol{j}^{2}_{t}\nonumber \\
&&+C^{\nabla j}_{t}\boldsymbol{s}_{t} \cdot\left(\boldsymbol{\nabla}\times\boldsymbol{j}_{t}\right)\nonumber \\
&&+C^{\nabla s}_{t} (\boldsymbol{\nabla}\cdot\boldsymbol{s}_{t})^2
+ C^{F}_{t} \boldsymbol{s}_{t} \cdot\boldsymbol{F}_{t},
\label{eq:5}
\end{eqnarray}
contains all time-odd densities and their derivatives written in a bilinear form.
The terms proportional to the coupling constants $C^{\nabla s}_{t}$
and $C^{F}_{t}$ occur for tensor force only and both are equal zero
in the standard parametrizations of the Skyrme effective interactions.

Invariance under local gauge transformations of the Skyrme energy density (\ref{eq:3})
links pairs of time-even and time-odd terms in the energy functional provided that
the coupling constants fulfill the constraints \cite{Doba95}:
\begin{align}
C^{\tau}_{t} &= -C^{j}_{t},\nonumber\\
C^{J0}_{t} &= -{\textstyle\frac{1}{3}}C^{T}_{t}- {\textstyle\frac{2}{3}}C^{F}_{t},\nonumber\\
C^{J1}_{t} &= -{\textstyle\frac{1}{2}}C^{T}_{t}+ {\textstyle\frac{1}{4}}C^{F}_{t},\\
C^{J2}_{t} &= -C^{T}_{t}- {\textstyle\frac{1}{2}}C^{F}_{t},\nonumber\\
C^{\nabla J}_{t} &= C^{\nabla j}_{t}.\nonumber
\end{align}
The spin-orbit terms are proportional only to $C^{\nabla J}_{t}$=$ C^{\nabla j}_{t}$
in the standard Skyrme functionals. However, with the generalized spin-orbit interaction
(with the full isovector freedom in the spin-orbit term \citep{Rein95})
\begin{align}
C^{\nabla J}_{0} &= -b -{\textstyle\frac{1}{2}}b',\nonumber\\
C^{\nabla J}_{1} &=    -{\textstyle\frac{1}{2}}b',
\end{align}
where $b$ and $b'$ are the new parameters.

Four zero-order coupling constants of the Skyrme EDF ($C^{\rho}_{t}[\rho_{0}]$,
$C^{s}_{t}[\rho_{0}]$) can be expressed in terms of the Skyrme force parameters
\cite{Skyrme} ($t_{0}$, $x_{0}$, $t_{3}$, $x_{3}$, $\alpha$) and the rest
(24 second-order) coupling constants can be expressed in terms of the other seven
Skyrme force parameters ($t_{1}$, $x_{1}$, $t_{2}$, $x_{2}$,
$W_{0}$, $t_{e}$, $t_{o}$), and therefore, the time-odd coupling constants in
the Skyrme EDF are linear combination of the time-even ones \citep{Doba95},
see also Ref.~\citep{Perl04,Bend09,Hell12} for further discussion.

The total energy in the Skyrme-HFB approach is
\begin{eqnarray}
E^{tot}[\boldsymbol{\bar{\rho}}] &\equiv& E^{tot}\left[\rho, \tau, \mathbb{J};
\boldsymbol{s}, \boldsymbol{T}, \boldsymbol{j}, \boldsymbol{F}; \tilde{\rho}\right]\nonumber \\
&=&
\int \mathrm{d}^{3} \boldsymbol{r} \left(\mathcal{E}_{kin}(\boldsymbol{r})+
\mathcal{E}_{Sk}(\boldsymbol{r})\right)\nonumber \\
&&+
\int \mathrm{d}^{3} \boldsymbol{r} \left(\mathcal{E}_{Coul}^{dir}(\boldsymbol{r})+
\mathcal{E}_{Coul}^{ex}(\boldsymbol{r})\right)\nonumber \\
&&+
\int \mathrm{d}^{3} \boldsymbol{r}\mathcal{E}_{pair}(\boldsymbol{r})+ E_{corr},
\label{eq:6}
\end{eqnarray}
where $\mathcal{E}_{kin}= \tau_{0}(\boldsymbol{r})({\hbar^{2}}/{2m})$
is a kinetic energy density of both protons and neutrons
(for the neutron and proton masses being approximated by their average value),
$\mathcal{E}_{Sk}$ is the Skyrme EDF, Eq.~(\ref{eq:3}),
and $\mathcal{E}_{Coul}^{dir}$, $\mathcal{E}_{Coul}^{ex}$ is a direct and an exchange
Coulomb energy density, respectively.

The $\mathcal{E}_{pair}$ is the isovector $|T_{z}| =$1 pairing energy density,
corresponding to a density-dependent delta interaction
\begin{equation}
\mathcal{E}_{pair}= \sum_{q=p,n}\frac{V^{0}_{q}}{4} \left[1-V^{1}
\left(\frac{\rho_{0}(\boldsymbol{r})}{\rho_{st}}\right)^{\beta} \right]
\tilde{\rho}^{2}_{q}(\boldsymbol{r}),
\label{eq:10}
\end{equation}
where $\rho_{st}$ is the saturation density of nuclear matter that
approaches the density inside the nucleus, $\beta=1$ (usually), and
$V^{1}=0,\, 1,\, \textrm{or}\, 1/2$ for \emph{volume-}, \emph{surface-},
or \emph{mix-}type pairing, and $\tilde{\rho}_{q}(\boldsymbol{r})$
is the paring density for protons and neutrons \cite{Doba84}.
The volume pairing interaction acts primarily inside the nuclear volume,
while the surface pairing acts on the nuclear surface.
A correction term, $E_{corr}$, includes corrections for spurious motions
caused by symmetry violation in the mean-field approximation \cite{Bend03}.

\subsection{The method of Lagrange multipliers}
The constrained and/or cranked Skyrme-HF(B) approach is equivalent to
minimization of the $E^{tot}$ EDF, Eq.~(\ref{eq:6}), with respect to
the densities and currents. Using the method of Lagrange multipliers we
solve an equality-constrained problem (ECP) for the objective function
$E^{tot}$:
\begin{equation}
\left\{
\begin{array}{l}
\displaystyle\min_{\boldsymbol{\bar{\rho}}} E^{tot}[\boldsymbol{\bar{\rho}}]\\
         \mbox{subject to: } \displaystyle\langle \hat{N}_{q} \rangle= N_{q},\quad (q=p,n),\\
\phantom{\mbox{subject to: }}\displaystyle\langle \hat{Q}_{\lambda\mu} \rangle= Q_{\lambda\mu},\\
\phantom{\mbox{subject to: }}\displaystyle\langle \hat{J}_{i} \rangle= I_{i},\quad (i=x,y,z),
\end{array}
\right. 
\label{eq:11}
\end{equation}
where the constraints are defined by average values $N_{p/n}=Z$ or $N$ for the proton and neutron
particle-number operator $\hat{N}_{p/n}$, the constrained values $Q_{\lambda\mu}$
for the mass-multiple-moment operators $\hat{Q}_{\lambda\mu}$, and the constrained value
$I_{i}$ for the angular momentum operator $\hat{J}_{i}$ along the $i$-axes.

To solve the above ECP one can use the standard method of Lagrange multipliers,
\textit{e.g.}, the quadratic penalty method, or the augmented Lagrangian method.
A comparison of both methods can be found in Ref.~\cite{alm}.

The augmented Lagrangian functional (or Routhian) associated with ECP is defined as
\begin{eqnarray}
E^{'}_{c}[\boldsymbol{\bar{\rho}},\boldsymbol{\lambda}, \boldsymbol{\Lambda}, \boldsymbol{\omega}]
&=& E^{tot}[\boldsymbol{\bar{\rho}}] -\sum_{q=p,n}\lambda_{q}\langle \hat{N}_{q}\rangle\nonumber \\
&&+\sum_{\lambda\mu}C_{\lambda\mu}\!\left(\langle \hat{Q}_{\lambda\mu} \rangle - Q_{\lambda\mu}\right)^{2}\nonumber \\
&&+\sum_{\lambda\mu}\Lambda_{\lambda\mu}\!\left(\langle \hat{Q}_{\lambda\mu} \rangle - Q_{\lambda\mu}\right)\nonumber \\
&&-\sum_{i=x,y,z}\omega_{i}\langle \hat{J}_{i}\rangle
\label{eq:15}
\end{eqnarray}
where $\lambda_{p}$, $\lambda_{n}$, $\Lambda_{\lambda\mu}$, and $\omega_{i}$ are
the \emph{Lagrange multipliers}, and $C_{\lambda\mu}>$0 are the \emph{penalty parameters}.
In the ALM the Lagrange multipliers $\Lambda_{\lambda\mu}$ are iterated according to
\begin{equation}
\Lambda_{\lambda\mu}^{k+1}= \Lambda_{\lambda\mu}^{k}+
2C_{\lambda\mu}^{k} \left(\langle \hat{Q}_{\lambda\mu} \rangle - Q_{\lambda\mu} \right),
\label{eq:15a}
\end{equation}
see, Ref.~\cite{alm} and references cited therein.

In an adiabatic approximation nuclear collective and intrinsic degrees of
freedom can be decoupled and the collective motion of nucleus can be described
in terms of a few collective variables describing shape evolution.
Using a primal function of ECP
\begin{equation}
E^{tot}(Q_{\lambda\mu};\boldsymbol{I})= \min_{\langle \hat{Q}_{\lambda\mu} \rangle
=Q_{\lambda\mu},\,\langle \hat{J}_{i} \rangle =I_{i}}
E^{tot}[\boldsymbol{\bar{\rho}}],
\label{eq:14a}
\end{equation}
one can characterize these shapes by the mean values of external fields represented
by the multipole-moments and angular momentum operators.

\subsection{The Skyrme-HFB calculations}
The Hartree-Fock wave function is the Slater determinant of single particle
orbitals. Thus the orbitals depend on the single particle Hamiltonian $\hat{h}$,
which depends on the densities and currents. The densities and currents in
turn depend on the orbitals, so we must solve ECP, Eq.~(\ref{eq:11}),
self-consistently (by iteration until convergence).

The above ECP was solved using the augmented Lagrangian method with the
symmetry-unrestricted code HFODD \cite{hfodd} which solves the Skyrme-HFB
equations in the Cartesian deformed harmonic-oscillator (h.o.) basis.
In the particle-hole channel the Skyrme SkM* force \cite{Bar82} was applied
and a density-dependent \textit{mixed} pairing \cite{Dob02,Sta09} interaction
with the parameters $V^{0}_{n}= -268.9$ MeV~fm$^3$ and $V^{0}_{p}= -332.5$
MeV~fm$^3$ in the particle-particle channel was used.

The code HFODD calculates parameters of the h.o. basis using geometrical
consideration \cite{hfodd12}. The relative values of the frequencies of
the deformed h.o. in the three Cartesian directions are defined by the
condition $\omega_{x}R_{x}=\omega_{y}R_{y}=\omega_{z}R_{z}$, while the
overall factor is given by $(\omega_{x}\omega_{y}\omega_{z})^{1/3}=\omega_{0}$,
where $\hbar \omega_{0}=f \times 41$ MeV$/A^{1/3}$ is the spherical h.o.
frequency and $f$=$1.2$ is a scaling factor \cite{hfodd12}.
In the above condition, $R_{x}$=$R(\pi/2,0)$, $R_{y}$=$R(\pi/2,\pi/2)$, and
$R_{z}$=$R(0,0)$ are the lengths of principal axes of a sharp-edge reference
body surface, defined by deformation parameters $\alpha_{\lambda\mu}$
in terms of multipole expansion
\begin{equation}
R(\theta, \phi)=c(\alpha)\Big( 1+ \sum_{\lambda=0}^{\lambda_{max}}
\sum_{\mu=-\lambda}^{\lambda} \alpha_{\lambda\mu}Y_{\lambda\mu}(\theta ,\phi)\Big),
\label{eq:12s}
\end{equation}
where $c(\alpha)$ is a function of $\alpha_{\lambda\mu}$ such that
the volume enclosed by the surface does not depend on $\alpha$.
In the present study, we have used the axially deformed h.o. basis with
the deformation parameter $\alpha_{20}$ chosen to be equal to the mean-field value
calculated in the code for a given value of $\langle \hat{Q}_{20} \rangle$,
\textit{cf.} Eq.~(1.35) of Ref.~\cite{Ring80} and Ref.~\cite{hfodd6}.
For example, this procedure for the quadrupole moment constraint $Q_{20}=-200$~b
gives $\alpha_{20}=-0.70$ which corresponds to
$\hbar\omega_{\perp}$=$\hbar\omega_{x}$=$\hbar\omega_{y}=5.96$ MeV and
$\hbar\omega_{z}=11.03$ MeV. We keep this deformed h.o. basis when we examine
toroidal shapes with the large oblate deformation $Q_{20}< -200$~b.
The basis was composed of the $1140$ lowest states taken from the $N_{0}$=26
h.o. shells. With this basis size, our tests show that we can properly describe
toroidal shapes up to $Q_{20}\gtrsim -600$~b deformation.

Our objective is to locate local toroidal figures of equilibrium, if any,
in the multi-dimensional search space of $\{A,Q_{20},I\}$.
We first use the quadrupole moment $Q_{20}$ constrained Skyrme-HFB approach
to search for the nuclear density distributions with toroidal shapes.
Next, using as starting configurations the toroidal solutions we apply
the constrained and cranked around the symmetry $z$-axis Skyrme-HF approach
to map out the energy landscape for axially-symmetric toroidal shapes
under $Q_{20}$ and $I$=$I_{z}$ constraints. If the states with
$I$=$I_{z}$ as a function of $Q_{20}$ deformation reveal a local energy
minimum then the quadrupole constraint is removed at that minimum
and symmetry-unrestricted free-convergence is tested to ensure that
the non-collectively rotating toroid nucleus is indeed a figure of equilibrium.
It is worth noting that in the unconstrained and symmetry-unrestricted
cranked Skyrme-HF calculations we do not impose the axial and reflection
symmetries to the toroidal system to ensure its stability with respect
to these modes.

\subsection{\label{pairing} Pairing correlations}
As mentioned above, in the present calculations we use the constrained
Skyrme-HFB approach only during the first stage of our method, when we
try to establish the region of $Q_{20}$ deformation with the toroidal
solutions. In the following calculations we apply the cranked Skyrme-HF
model (neglecting the pairing correlations) trying to locate the THSIs.

A quantal system such as axially symmetric toroid cannot rotate around
a symmetry axis. In the cranking approach the Lagrangian multiplier
$\omega_{z}$ allows one to solve the ECP (\ref{eq:11}) with a supplementary
condition on an angular momentum $\langle \hat{J}_{z} \rangle =I_{z}$,
where the $z$-axis is choose as the symmetry axis.
The total angular momentum $I$=$I_{z}$, in a case when $\omega_{x}$=
$\omega_{y}$=0, is built up by selecting nucleonic orbitals that are
most favorable in creating the states with required angular momentum
and with the lowest energy, the so-called optimal configurations
(\textit{cf.} Refs.~\cite{And76,Boh81,Ring80,Voi83,Nils95,Afa99}).
This non-collective rotation around the symmetry axis is permissible
quantum mechanically only by particle-hole excitations with respect to the uncranked
state, leading to aligned single-particle angular momenta along the
symmetry axis
\begin{eqnarray}
I_{z}=\langle \hat{J}_{z} \rangle &=& \sum_{i=1}^{A} \langle \hat{j}_{z} \rangle_{i}
=\sum_{i}^{A} (\Omega_{z})_{i}\nonumber\\
&=& \sum_{i\,\, exc} \left(\Omega^{part}_{z}- \Omega^{hole}_{z}\right)_{i},
\label{eq:16}
\end{eqnarray}
where $\Omega_{z}=\Lambda_{z}\pm 1/2$ denotes the projection of the single-particle
angular momentum onto the symmetry $z$-axis and in the second equation the sum runs over
the particle-hole excitations.

The Cooper pairs in  a nucleus are composed of the pairs of nucleons in the
time-reversal conjugate orbitals with $\Omega_{z}=\pm \Omega$.
The pairing correlation diminishes with each particle-hole excitation
which successively breaks down the Cooper pairs. When the seniority of
a configuration increases, the blocking effect \cite{Ring80,Afa99} is effective in
reducing the pairing correlations in the toroidal high-spin states. We neglect
the pairing in the present study of the THSIs. It would certainly be interesting
to examine the effect of weak pairing correlations on toroidal high-spin isomers,
but that will be left for a future study.

\section{Results and discussions}

\subsection{Toroidal system of $^{304}{120}_{184}$ without spin}

Using the above self-consistent Skyrme-HFB mean-field theory, we study first
the nucleus $^{304}120_{184}$ under the constraint of a fixed $Q_{20}$ without
spin. We obtain the total energy of the system  with a toroidal  density as
a function of the constrained $Q_{20}$, as shown in Fig.~\ref{Fig2}.
It indicates that even though $^{304}120_{184}$ without spin may have a toroidal
density for $Q_{20}\le$ -158 b, its total energy curve as a function of $Q_{20}$
lies on a slope. This implies that the toroidal system of $^{304}120_{184}$
without spin is unstable against the tendency to return to a sphere-like
geometry, \textit{cf.} Ref.~\cite{Kosi17}. For future exploration of possible superheavy
toroidal nuclear system without spin, it will be necessary to go to systems with a greater
charge numbers with $Z\ge 122$ as in Ref.~\cite{Sta08} or alternatively to find
single-particle ``shells" in proton and neutron numbers in regions of sparse single-particle
level densities at the top of the Fermi surface, for which the shell effects may provide
a sufficiently shell correction \cite{Bra72} to stabilize a toroidal nuclear system.

\begin{figure}[htb]
\begin{center}
\includegraphics[width=\columnwidth]{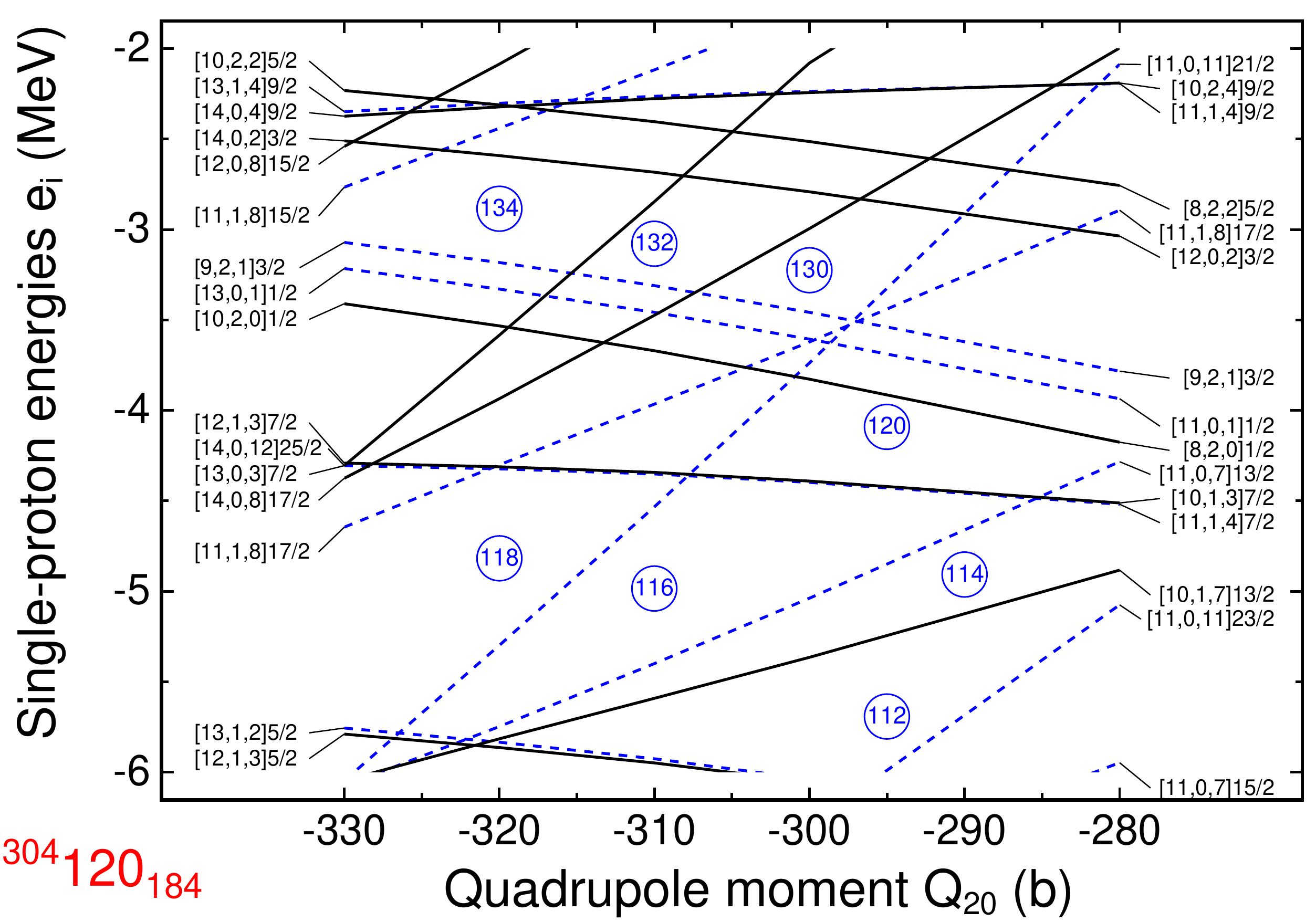}
  \caption{\label{Fig3} (Color online) Proton single-particle levels
  in the canonical basis for $^{304}120_{184}$ in the toroidal configuration
  as a function of the constraining quadrupole moment $Q_{20}$,
  obtained in the Skyrme-HFB calculations.
  The levels with positive parity are drawn with solid (black) lines, while
  those with negative parity are drawn with gray dashed (blue color) lines.
  The circled numbers denote the occupation numbers at regions of spare
  single-particle energy level density (``shells").}
\end{center}
\end{figure}

\begin{figure}[htb]
\begin{center}
\includegraphics[width=\columnwidth]{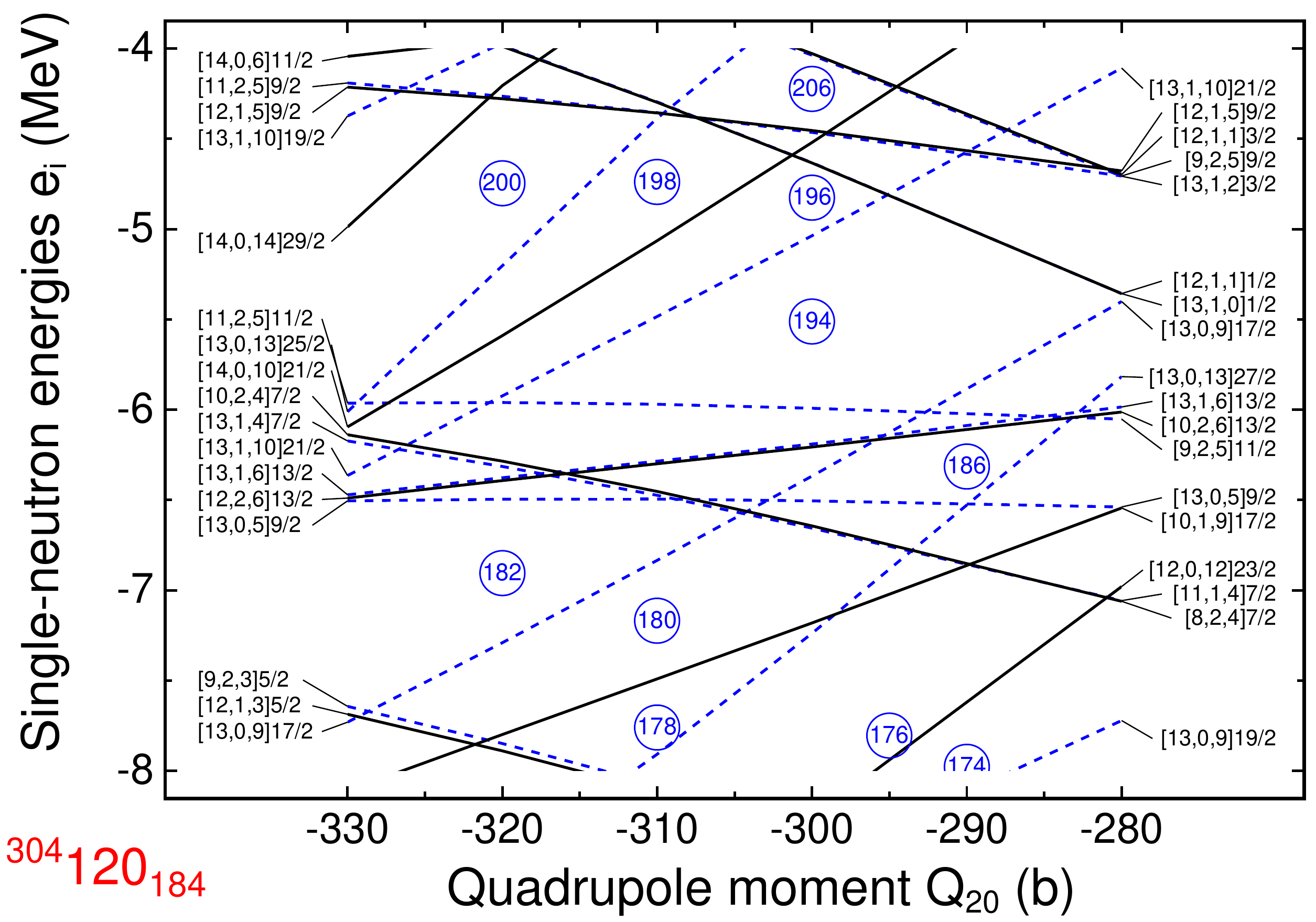}
  \caption{\label{Fig4} (Color online) The same as in Fig.~\ref{Fig3},
  but for the neutron single-particle levels.}
\end{center}
\end{figure}

To study the shell effects in superheavy toroidal nuclear system without spin,
we examine the single-particle states of $^{304}120_{184}$ with a toroidal density
as a function of the quadrupole moment $Q_{20}$ in self-consistent Skyrme-HFB
calculations.
The self-consistent single-particle potential will also assume a toroidal shape.
The proton and neutron single-particle energy levels (in the canonical basis) for
$^{304}120_{184}$ are shown in Fig.~\ref{Fig3} and in Fig.~\ref{Fig4}, respectively.
Each single-particle state is labeled by the Nilsson quantum numbers
$[N,n_z,\Lambda] \Omega$ of the dominant component, and is twofold degenerate, with
$\Omega_z=\pm \Omega$. Solid and dashed curves are used to distinguish positive and
negative parity levels, respectively.
We find from Figs.~\ref{Fig3} and \ref{Fig4} that the densities of neutron and proton
single-particle states are far from uniform. There are regions of sparse  single-particle
level densities which can be identified as the ``shells" associated with enhanced
stability \cite{Bra72}.
For brevity of notation, we shall call these shells associated with a toridal nuclear
density and potential the {\it toroidal} shells.

For the nucleus $^{304}120_{184}$ in the toroidal configuration at
$Q_{20}\approx -300$~b, Figs.~\ref{Fig3} and \ref{Fig4} show that the proton Fermi
surface for $Z=120$ resides in the low single-particle level density region of
a proton shell at $Z=120$, but the neutron Fermi surface for $N=184$ resides
in a region of high single-particle level density.
The stabilizing effects for the \textit{toroidal} proton shell at $Z=120$ with a negative
proton shell correction is counter-balanced by the destabilizing effect for $N=184$
with a positive neutron shell correction, in the region of deformation
$Q_{20}\approx -300$~b.
Furthermore, the bulk Coulomb interaction in $^{304}120_{184}$ nucleus is just below
the threshold to open up a hole for a toroidal system, as it is for a nucleus
with $Z\ge 122$ \cite{Sta08}. As a consequence, in the Skyrme-HFB approach which
takes into account both the bulk properties and the shell effects, the combined total
energy of $^{304}120_{184}$ without spin in the toroidal configuration does not
possess an energy minimum as a function of $Q_{20}$.

Even though Figs.~\ref{Fig3} and \ref{Fig4} pertain to the self-consistent single-particle
states for $^{304}120_{184}$, we expect that as the mean-field potential varies only slightly
as a function of the atomic number and the neutrons number, and it depends more sensitively
on the spatial shape of the nuclear density distribution, the single-particle state diagrams
in Figs.~\ref{Fig3} and \ref{Fig4} can therefore be approximately applied as single-particle
states for the deformations $Q_{20}$ in the toroidal configuration in an extended region
around $^{304}120_{184}$.
One can therefore read out various \textit{toroidal} shells for protons and neutrons at various
deformations $Q_{20}$ in Figs.~\ref{Fig3} and \ref{Fig4}. One finds proton
shells at $Z$=116, 118, 120, 132, 134, and neutron shells at $N$=180, 182, 186, 194,
and 198. In our future work, we will exploit the property of the extra stability of superheavy
nuclei for which the \textit{toroidal} proton and neutron shells are located at the same
deformation.

\subsection{Construction of toroidal configurations of $^{304}120_{184}$ with high spin}

As the toroidal configurations of $^{304}120_{184}$ nucleus are unstable without spin, we like
to examine here whether toroidal $^{304}120_{184}$ may be stabilized when it
possesses an angular momentum aligned along the symmetry axis such that $I$=$I_z$.
Following Bohr and Mottelson \cite{Boh81}, we can construct a nucleus with
an aligned angular momentum $I_z$ by particle-hole excitations.
Specifically, referring to the single-particle states in Figs.~\ref{Fig3} and \ref{Fig4}
for toroidal system of $^{304}120_{184}$ at $Q_{20}=-300$~b without spin, we can make
a hole at a state with angular momentum component $-|\Omega_{z}^{hole}|$ and
place it at a particle state with angular momentum $\Omega _{z}^{part}$.
The particle-hole pair will generate an aligned angular momentum $I_z$ of
magnitude $\Omega _{z}^{part}+|\Omega _{z}^{hole}|$, see, Eq.~(\ref{eq:16}).
By making many such particle-hole excitations, a nucleus with a very high spin,
$I$=$I_z$, can be constructed, especially when the number of particle-hole
excitations and the magnitudes $|\Omega_z|$ of these participating particle
and hole states are large. Because $I_z$ depends on $\Omega_z$ and the number
of particle-hole excitations, it assumes quantized non-trivial values that can
only be obtained from a detailed examination of the structure of the single-particle
state energy diagram of the nucleus of interest.

There are two equivalent ways to construct a high-spin state with the spin
aligned along the symmetry axis: (i) the method of employing the tilted Fermi surfaces,
and (ii) the plots of the single-particle Routhians $e'_{i}=e_{i}-\hbar \omega (\Omega_{z})_{i}$
as a function of $\hbar \omega$.

\begin{figure}[htb]
\begin{center}
\includegraphics[width=\columnwidth]{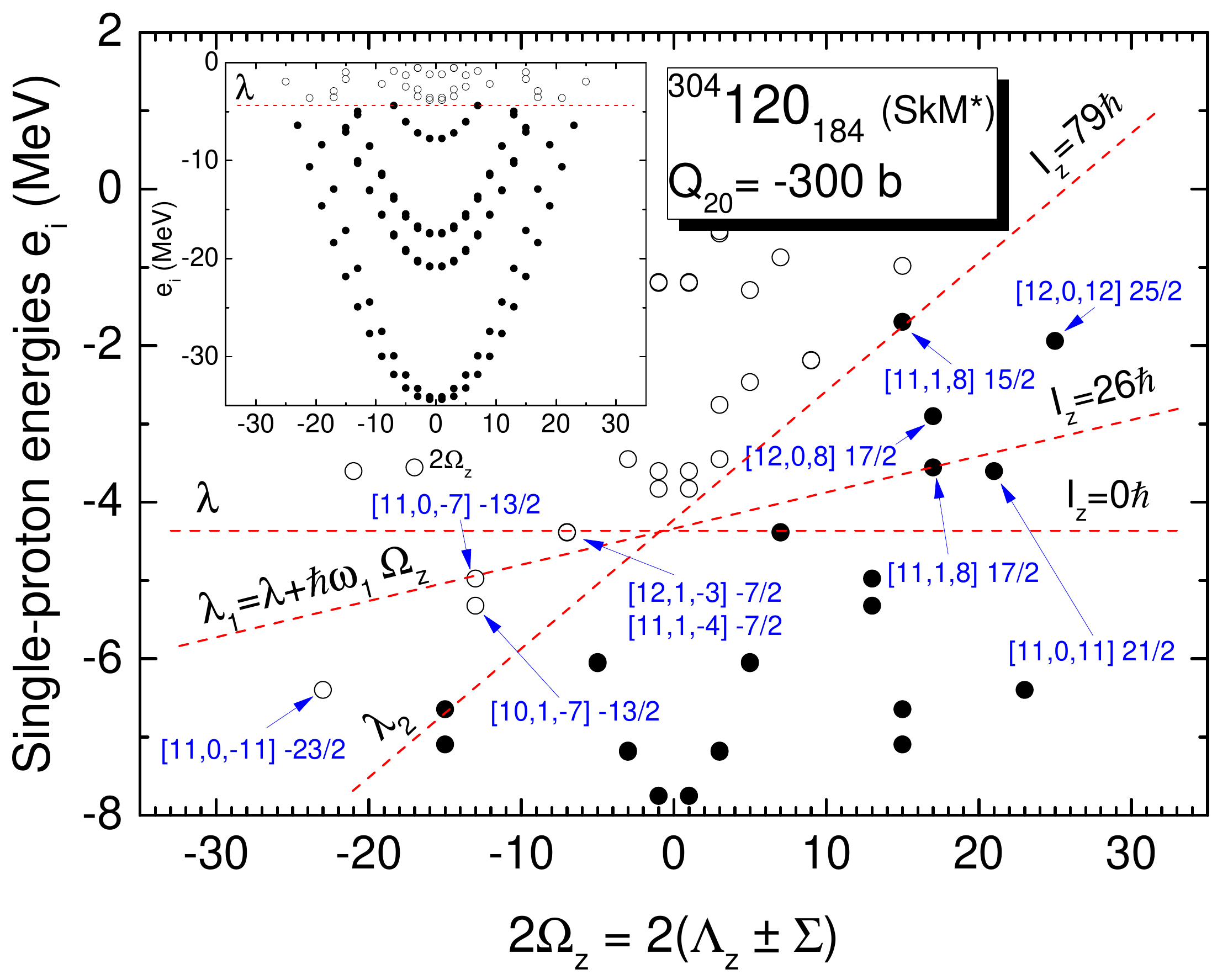}
  \caption{\label{Fig5} (Color online) The proton single-particle energy
  levels  of $^{304}120_{184}$ in the toroidal configuration at $Q_{20}=-300$~b, as a function
  of $2 \Omega_z$. The thin gray dashed (red color) lines give the tilted proton Fermi surfaces which
  lead to the proton spin value $I_z$=26 for $\hbar \omega_{1}$$\approx$0.1 MeV, and $I_z$=79
  at $\hbar \omega_{2}$$\approx$0.28 MeV. In the case of $I_z$=79, the occupied states are shown
  as solid circular points, and the unoccupied states as open circles.}
\end{center}
\end{figure}

\begin{figure}[htb]
\begin{center}
\includegraphics[width=\columnwidth]{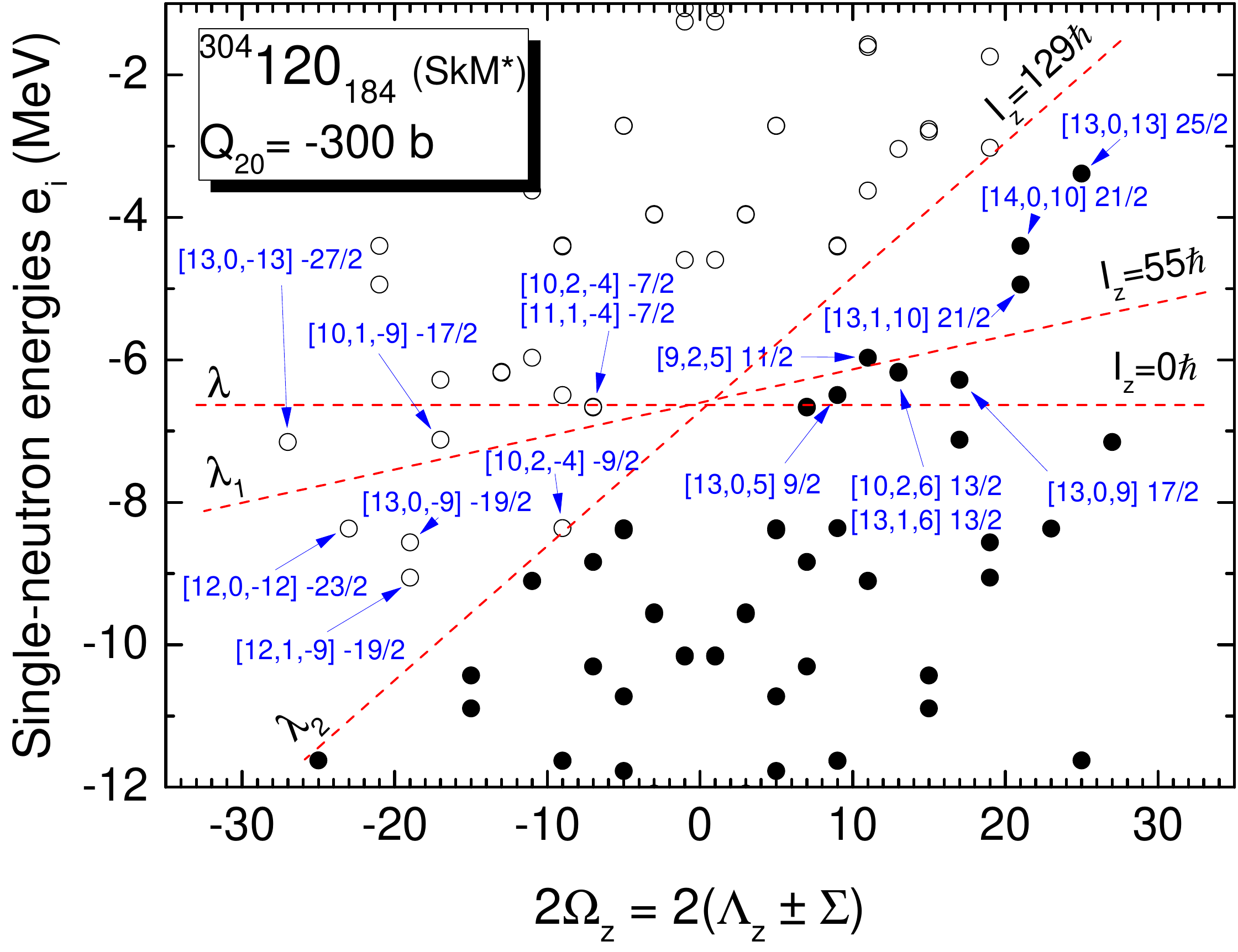}
  \caption{\label{Fig6}(Color online)  The neutron single-particle energy levels of $^{304}120_{184}$
  in the toridal configuration at $Q_{20}=-300$~b, as a function of $2 \Omega_z$.
  The thin dashed lines give the tilted neutron Fermi surfaces which
  lead to the neutron spin value $I_z$=55 for $\hbar \omega_{1}$$\approx$0.1 MeV, and $I_z$=129
  for $\hbar \omega_{2}$$\approx$0.28 MeV. In the case of $I_z$=129, the occupied states are shown
  as solid circular points, and the unoccupied states as open circles.}
\end{center}
\end{figure}

The single-particle energy level diagram at a fixed quadrupole moment, say
$Q_{2}=-300$~b, can be expanded out to include the additional dependence of
$\Omega_z$ as the horizontal axis, as shown in Figs.~\ref{Fig5} and \ref{Fig6}.
The Fermi surface for this case without spin shows up as a horizontal line
and all levels below it are occupied, see, an inset in Fig.~\ref{Fig5}.
A high-spin state can be constructed by tilting the Fermi level in the expanded
single-particle diagram, \textit{cf.} Ref.~\cite{Voi83}. The degree of tilt
can be specified in the Skyrme-HF calculations by the Lagrange multiplier
$\hbar \omega$ which describes the constraint $I_{z}$=$\langle \hat{J}_{z} \rangle$=
$\sum_{i=1}^{A} (\Omega_{z})_{i}$, with each $I$=$I_z$ spanning a small region of
$\hbar \omega$ \cite{Ring80}.

We collect in Table~\ref{Tab1} the particle-hole excitation configurations leading
to the states of $^{304}120_{184}$ with $I_z=81$ and 208. They are particle-hole
excitations relative to the Skyrme-HFB states without spin, as labeled by the
quantum numbers $[N,n_{z},\Lambda_{z}]\Omega_{z}$ for the optimal toroidal
configurations of $^{304}120_{184}$ at $Q_{20}=-300$~b in Figs.~\ref{Fig5} and
\ref{Fig6}.

\begin{table}[htb]
\caption {\label{Tab1} The particle-hole excitation configurations leading to the
states of $^{304}120_{184}$ with $I_z$=$I_z($proton) +$I_z($neutron)=26+55=81 and
$I_z$=79+129=208.}
\begin{ruledtabular}
\begin{tabular}{lll}
                 & Hole states &Particle states\\
\hline
                 &[11,1,-4] -7/2&[11,0,11] 21/2\\
$I_z$(proton)=26 &[12,1,-3] -7/2&[11,1,8] 17/2\\
\hline
                 &[11,0,-7] -13/2&[12,0,8] 17/2\\
                 &[10,1,-7] -13/2&[12,0,12] 25/2\\
$I_z$(proton)=79 &[11,0,-11] -23/2&[11,1,8] 15/2\\
\hline
\hline
                   &[10,2,-4] -7/2&[13,0,5] 9/2\\
                   &[11,1,-4] -7/2&[13,0,9] 17/2\\
                   &[10,1,-9] -17/2&[13,1,6] 13/2\\
$I_z$(neutron)=55  &[13,0,-13] -27/2&[10,2,6] 13/2\\
\hline
                   &[12,0,-12] -23/2&[9,2,5] 11/2\\
                   &[13,0,-9] -19/2&[13,1,10] 21/2\\
                   &[12,1,-9] -19/2&[14,0,10] 21/2\\
$I_z$(neutron)=129 &[10,2,-4] -9/2&[13,0,13] 25/2\\
\end{tabular}
\end{ruledtabular}
\end{table}

\begin{figure}[htb]
\begin{center}
\includegraphics[width=\columnwidth]{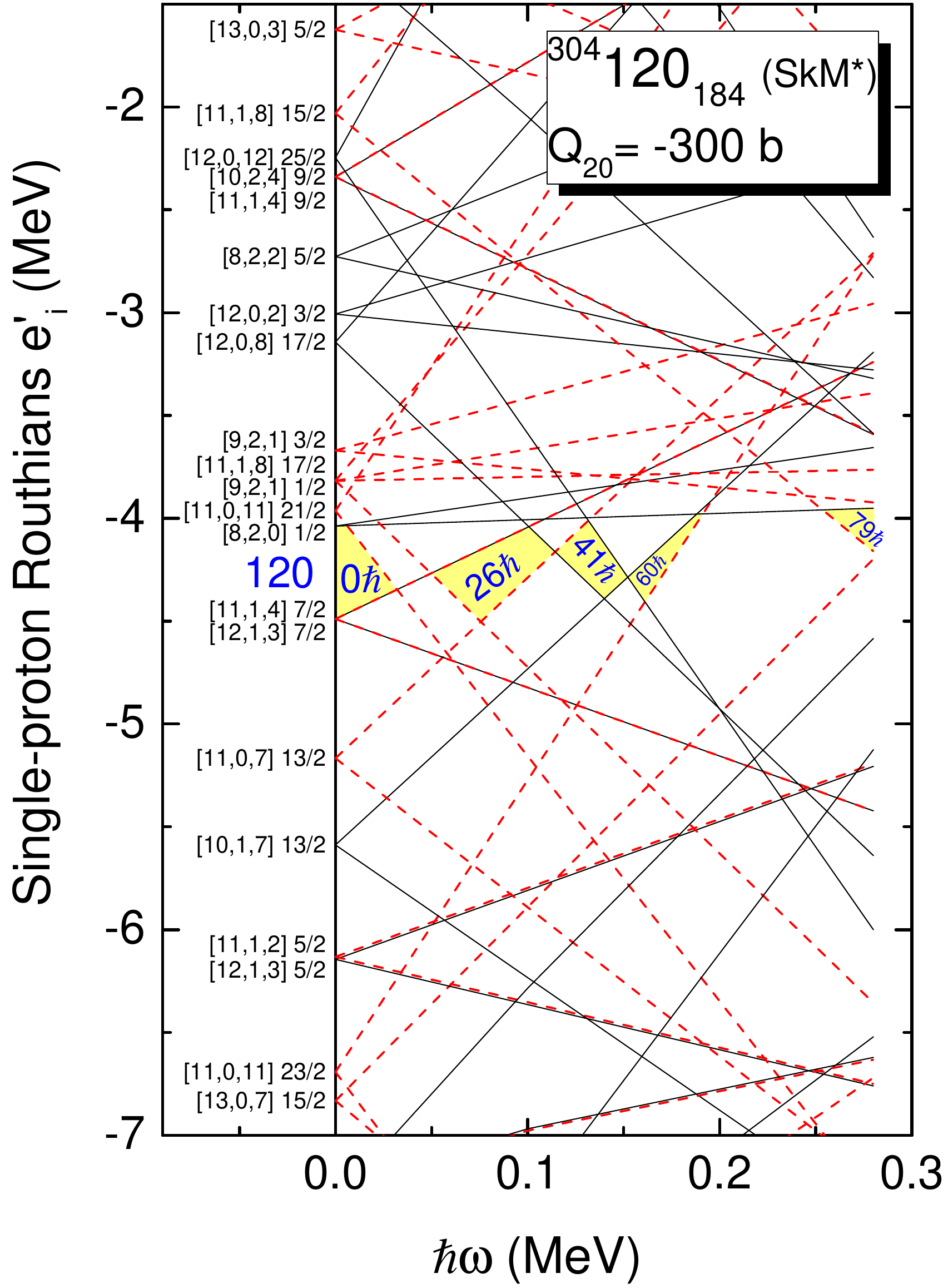}
  \caption{\label{Fig7} (Color online) Proton single-particle Routhians of
  $^{304}{120}_{184}$ in the toroidal configuration with $Q_{20}$=$-$300 b, 
  as a function of the cranking frequency $\hbar \omega$. The states are labeled 
  by the Nilsson quantum numbers $[N,n_{z},\Lambda] \Omega$.
  Solid (black) and dark gray dashed (red color) curves are used to distinguish 
  even and odd principal quantum number states, respectively. The aligned angular 
  momenta $I_z$ for $Z=120$ protons are shown at various $ \hbar \omega$ locations.}
\end{center}
\end{figure}

\begin{figure}[htb]
\begin{center}
\includegraphics[width=\columnwidth]{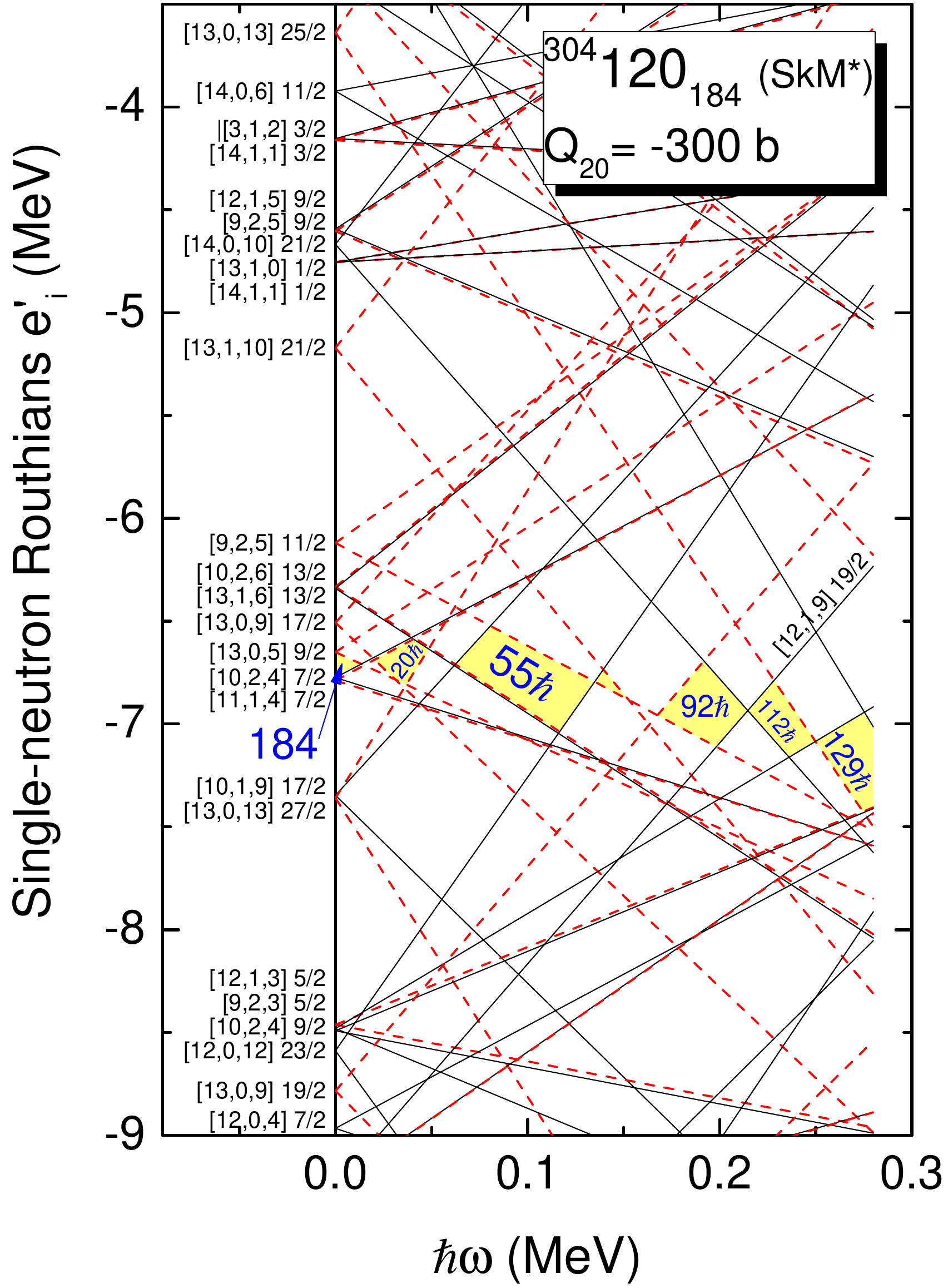}
  \caption{\label{Fig8} (Color online) The same as in Fig.~\ref{Fig7},
  but for the neutron single-particle Routhians of  $^{304}120_{184}$ in the toroidal configuration.
  The aligned angular momenta $I_z$ for $N=184$ neutrons are shown at various $\hbar \omega$
  locations.}
\end{center}
\end{figure}

In addition to the tilted Fermi surface method, there is another equivalent method
using the diagrams of single-particle Routhians vs. $\hbar \omega$.
Upon using a Lagrange multiplier $\hbar \omega$ to describe the constraint of an
aligned angular momentum $I$=$I_z$ along the symmetry $z$-axis, the constrained
single-particle Hamiltonian becomes $\hat{h}'={\hat h}-\hbar \omega \hat{j}_{z}$,
where $\hat{j}_{z}$ is the $z$-component  of the single-particle angular momentum
operator along the symmetry axis with eigenvalue $\Omega_z$. The single-particle
Routhian $e_i'$ is the eigenvalues of $\hat{h}'$ for the single-particle state $i$.
A nucleus in the state with a total aligned angular momentum $I_z$ along the symmetry
axis can be constructed by populating states below the Fermi level in the single-particle
Routhian level diagram.
As the Routhian $e'_{i} (\hbar \omega) $ for the state $\Omega_z$ is shifted
from the corresponding single-particle energy without spin $e'_{i} |_{\hbar \omega=0}$
by a term proportional to $-\hbar \omega (\Omega_{z})_{i}$, different Lagrange multipliers
$\hbar \omega$ will result in different ordering of the single-particle Routhians
and different $I_z$, for a given occupation number $Z$ or $N$.
In Figs.~\ref{Fig7} and \ref{Fig8}, we give the proton and neutron single-particle
Routhians as a function of the constraining Lagrange multiplier $\hbar \omega$,
for a toroidal system of $^{304}120_{184}$ with $Q_{20}=-300$ b, obtained in
self-consistent cranked Skyrme-HF calculations.

We can use single-particle Routhians in Figs.~\ref{Fig7} and \ref{Fig8} to determine
$I_z$ as a function of the nucleon occupation number $N_{p/n}$ and $\hbar \omega$.
For a given $N_{p/n}$ and $\hbar \omega$, the aligned $I_z$ angular momentum can be
obtained by summing $\Omega_{z}$ over all states below the Fermi surface, \textit{cf.}
Eq. (\ref{eq:16}). For the occupation numbers of $Z=120$ and $N=184$ in Figs.~\ref{Fig7}
and \ref{Fig8}, there are shells, regions of low Routhian energy level density, for
different $I_{z}$ configurations at different values of $\hbar \omega$. They represent
configurations with relative enhanced stability \cite{Bra72,Won73}. In the corresponding
Skyrme-HF calculation, they may lead to local energy minima for various allowed
angular momenta.

Figure~\ref{Fig7} shows that for the proton occupation number $Z=120$, possible
shells are located at $I_z$(proton)=0, 26, 41, 60, and 79 at different values
of $\hbar \omega$. Figure~\ref{Fig8} shows that for the neutron occupation number
$N=184$, possible shells are $I_z$(neutron)=0, 20, 55, 92, 112, and 129 at various
values of $\hbar \omega$. For a nucleus to have a local minimum with a total aligned
angular momentum $I_z$=$I_z({\rm proton})$+$I_z({\rm neutron}) $, the $\hbar \omega$
locations of the proton and neutron shells, need to be close to each other. We find
that by combining the proton and neutron spins, the total spin of the system can be
$I_z$= 81 at $\hbar \omega$$\approx$0.1~MeV, and $I_z$=208 at $\hbar \omega$$\approx$0.28~MeV,
for $Q_{20}=-300$~b.

Referring to the proton single-particle Routhians diagram at $\hbar \omega$$\approx$0.1~MeV
in Fig.~\ref{Fig7}, the proton spin of $I_z({\rm proton})$=26 for the 2p-2h
excitation arises by emptying the \mbox{[11,1,-4]-7/2} and \mbox{[12,1,-3]-7/2}
states, and occupying \mbox{[11,0,11]21/2} and \mbox{[11,1,8]17/2} states.
This result in the alignment of $I_z$=7 from the holes, $I_z$=19 from particles,
and $I_z({\rm proton})$=7+19=26, \textit{cf.} Eq. (\ref{eq:16}).
In  Fig.~\ref{Fig8}, the neutron spin of $I_z({\rm  neutron})$=55 arises by
emptying \mbox{[10,2,-4]-7/2}, \mbox{[11,1,-4]-7/2}, \mbox{[10,1,-9]-17/2},
and \mbox{[13,0,-13]-27/2} states, and populating \mbox{[13,0,5]9/2}, \mbox{[13,0,9]17/2},
\mbox{[13,1,6]13/2}, and \mbox{[10,2,6]13/2} states. This results in $I_z$(neutron)=
29+26=55 for which the neutron holes provide 29 units, and the neutron particles 26.
The total spin of the toroidal system of $^{304}120_{184}$ at $\hbar \omega$$\approx$0.1~MeV
is $I_z$=$I_z($proton)+$I_z($neutron)=26+55=81.

For the nuclear total spin of $I_z$=208 at $\hbar \omega$$\approx$0.28~MeV, one
observes from Fig.~\ref{Fig7} that the proton spin of $I_z({\rm proton})$=79 from
the 5p-5h excitation arises by emptying the proton states \mbox{[11,1,-4]-7/2},
\mbox{ [12,1,-3]-7/2}, \mbox{[11,0,-7]-13/2}, \mbox{[10,1,-7]-13/2}, and
\mbox{[11,0,-11]-23/2}, and occupying proton states \mbox{[11,0,11]21/2},
\mbox{[11,1,8]17/2}, \mbox{[12,0,8]17/2}, \mbox{[12,0,12]25/2}, and \mbox{[11,1,8]15/2}.
This result in the alignment of $I_Z$=(63/2) from the holes, and $I_Z$=(95/2)
from particles. The proton 5p-5h excitation gives $I_z$(proton)=(63/2)+(95/2)=79.
In Fig.~\ref{Fig8}, the neutron spin of $I_z({\rm neutron}$)=129 arises from
the 8p-8h excitation by emptying \mbox{[10,2,-4]-7/2}, \mbox{[11,1,-4]-7/2},
\mbox{[10,1,-9]-17/2}, \mbox{[13,0,-13]-27/2}, \mbox{[12,0,-12]-23/2},
\mbox{[13,0,-9]-19/2}, \mbox{[12,1,-9]-19/2}, \mbox{[10,2,-4]-9/2} states,
and populating \mbox{[13,0,5]9/2}, \mbox{[13 ,0 ,9 ]17/2}, \mbox{[13,1,6]13/2},
\mbox{[10,2,6]13/2}, \mbox{[9,2,5]11/2}, \mbox{[13,1,10]21/2}, \mbox{[14,0,10]21/2},
and \mbox{[13,0,13]25/2} states.
The neutron 8p-8h excitation gives $I_z$(neutron)=64+65=129.

The self-consistent single-particle Hamiltonian $\hat{h}'$ under an aligned angular
momentum constraint depends on the Hamiltonian operator $\hat{h}$ that is a self-consistent
function of the nuclear density and nuclear current. The latter nuclear current depends
on the aligned angular momenta $I_z$, which depends in turn on the Lagrange multiplier
$\hbar \omega$. Therefore, the single-particle Routhian, $e'_{i}$, which is the eigenvalue
of $\hat{h}'$, can acquire an additional self-consistency dependence on $\hbar \omega$,
in addition to the explicit dependency on $-\hbar \omega \Omega_z$. We find that
the self-consistent Skyrme-HF single-particle Routhians
$e'_{N n_z \Lambda_{z} \Omega_{z}}({\hbar \omega})$ in Figs.~\ref{Fig7} and \ref{Fig8}
can be approximately represented by
\begin{equation}
e'_{ N n_z \Lambda_z  \Omega_z}(\hbar \omega)\approx
e'_{N n_z \Lambda_z \Omega_z}|_{\hbar \omega=0} + a \hbar \omega
-\hbar \omega \Omega_z,
\end{equation}
where the additional term $a \hbar \omega$ with a parameter $a\approx0.5$ arises from
the \textit{effect of self-consistency} of the single-particle Routhian Hamiltonian
$\hat{h}'$. It affects mostly those states with a small value of $\Omega_z$ and
is un-important for states with large $\Omega_z$. In the present case for proton
occupation number at $Z=120$, an $\Omega_z=$1/2 state occurs by chance at the top
to the Fermi surface, as in Fig.~\ref{Fig7}.

\subsection{The toroidal high-spin isomers in $^{304}120_{184}$}

The tilted Fermi surface method or the Routhian single-particle method in the last
subsection deals only with the construction of a state with an aligned angular
momentum along the toroidal symmetry axis. The question of the stability for such
a nucleus needs to be examined by studying the dependence of the total energy as
a function of  $Q_{20}$ and $I_z$.
The investigation can be carried out by extending the Skyrme-HFB calculations
further to include both the quadrupole moment $Q_{20}$ constraint and the angular
momentum constraint, $I$=$I_z$ using a Lagrange multiplier $\hbar \omega$
as the cranking frequency.
As stated in Sec.~\ref{pairing} we have carried out the cranking calculations without
the pairing interaction, using the cranked Skyrme-HF approach.

Applying an additional constraint of an angular momentum $I=I_{z}$ about
the symmetry $z$-axis in the cranked Skyrme-HF calculations, we search for
the energy minima of ${}^{304}120_{184}$ in the toroidal configuration as a function of the
deformation $Q_{20}$ and aligned angular momentum $I_z$. If a local energy
minimum with $I=I_{z}$ is found, we perform at this point the cranked
symmetry-unrestricted and deformation unconstrained Skyrme-HF calculations
to locate a stable THSI state in free convergence.

\begin{figure}[htb]
\begin{center}
\includegraphics[width=\columnwidth]{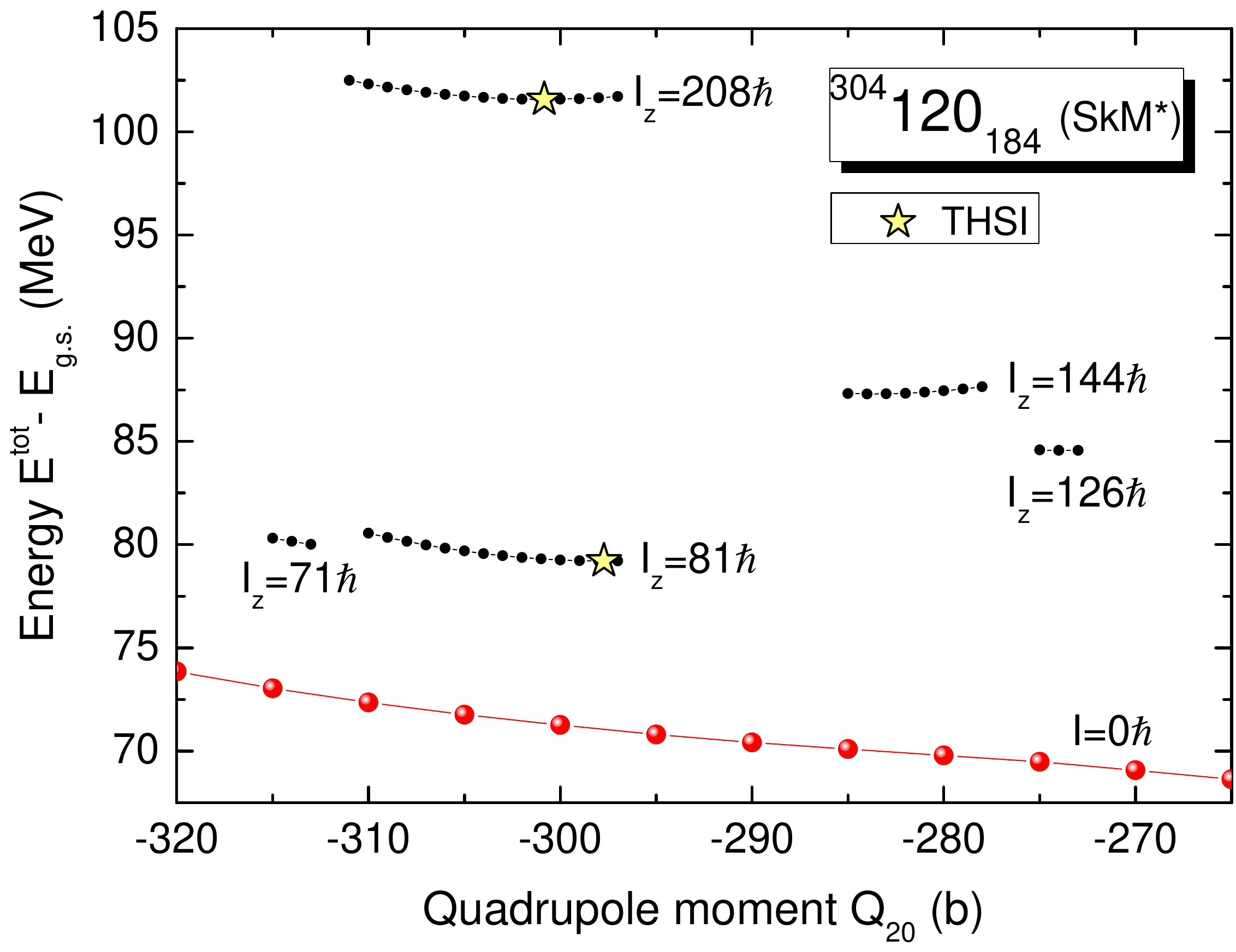}
  \caption{\label{Fig9} (Color online) The deformation energies of $^{304}120_{184}$
  in the toroidal configuration
  as a function of the quadrupole moment $Q_{20}$ for $I_z$=0, 71, 81, 126, 144, and 208.
  The locations of the toroidal high-spin-isomers (THSIs) for $I_z$=81 and 208 are indicated
  by star symbols. All deformation energies are measured relative to the energy of the spherical
  ground state.}
\end{center}
\end{figure}

The results of such calculations for $^{304}120_{184}$ are presented in
Fig.~\ref{Fig9}, where we plot the deformation energy (relative to the
spherical ground state energy) of the high-spin toroidal states as a function
of the constrained $Q_{20}$, for different quantized $I_z$. For each point
$( Q_{20},I_z)$ on an $I_z$ curve, it was necessary to adjust $\hbar \omega$
to ensure that the total aligned angular momentum of all nucleons in the
occupied states gives the quantized $I_z$ value of interest.

\begin{figure}[htb]
\begin{center}
\includegraphics[width=\columnwidth]{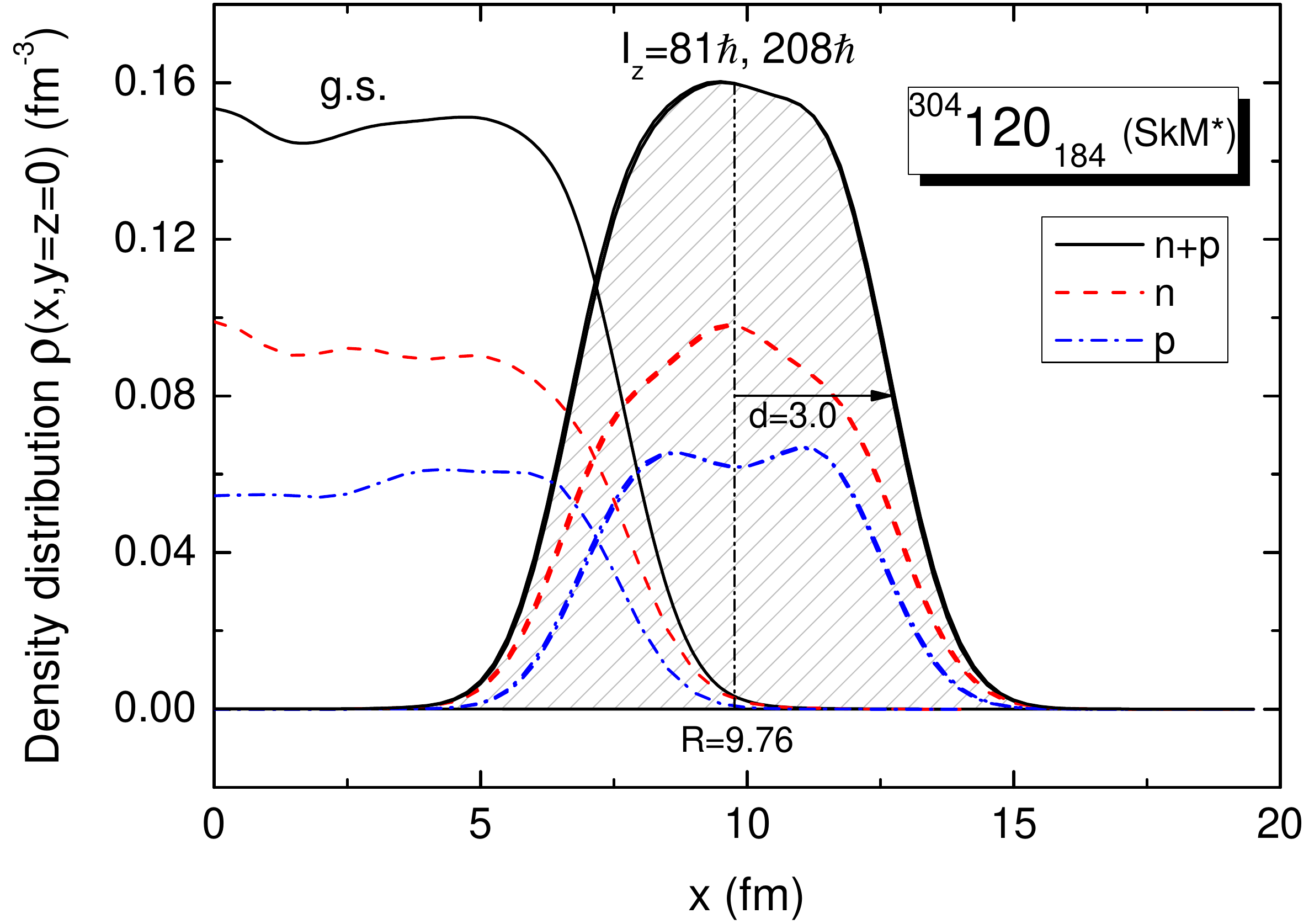}
  \caption{\label{Fig10} (Color online)  Neuron, proton and total density profiles
  of the THSIs $^{304}120_{184}$($I$=81 and 208) as a function of $x$ for a cut in
  $y=0$, and $z=0$.}
\end{center}
\end{figure}

From the energy surface of $^{304}120_{184}$($I_z$=81) in the toroidal configuration
in Fig.~\ref{Fig9}, we find that when we vary the constrained $Q_{20}$ with
$\hbar \omega$$\approx$0.1~MeV, the deformation energy of the nucleus in
the toroidal configuration as a function of $Q_{20}$ has a minimum.
Similarly, from the energy surface of $^{304}120_{184}$($I_z$=208),
we find that when we vary the constrained $Q_{20}$ with $\hbar \omega$$\approx$0.28~MeV,
the deformation energy of the nucleus as a function of $Q_{20}$ has a minimum.
Thus, we have theoretically located two THSI states of
$^{304}{120}_{184}$ with an angular momentum $I$=$I_z$=81 (proton 2p-2h, neutron
4p-4h excitation) and $I$=$I_z$=208 (proton 5p-5h, neutron 8p-8h) at $Q_{20}=-297.7$~b
and $Q_{20}=-300.8$~b  with energies 79.2 MeV and 101.6 MeV above the spherical
ground state energy, respectively.
In Fig.~\ref{Fig9}, deformation energies for  $I$=$I_z$=126 at $Q_{20}\sim -275$~b
and $I$=$I_z$=144 at $Q_{20}\sim -280$~b are also exhibited. As there are no energy
minima for these $I_z$ states, there are no toroidal high-spin isomers with these
aligned angular momenta.

\begin{figure}[htb]
\begin{center}
\includegraphics[width=0.8\columnwidth]{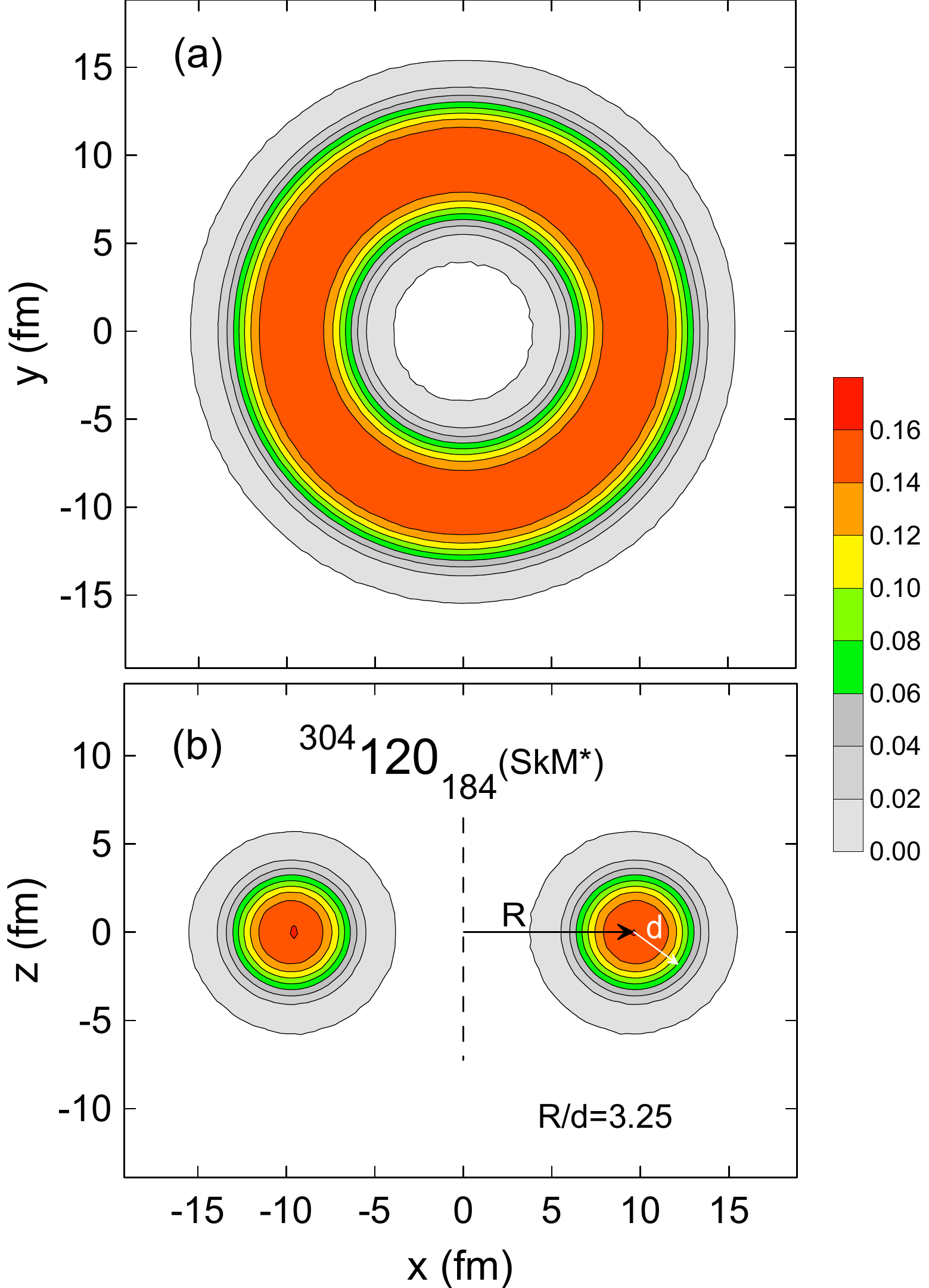}
  \caption{\label{Fig11} (Color online) Contours of the total nuclear densities of
  $^{304}120_{184}$($I_z$=81) in cuts: $x-y$ (a), and $x-z$ (b).}
\end{center}
\end{figure}

After the THSIs $^{204}120_{184}$ with $I$=$I_z$=(81 and 208) have been located,
we can examine their properties. Their density profiles as a cut in the plane of positive $x$
is shown in Fig.~\ref{Fig10}, and as density contours in Fig.~\ref{Fig11}. The corresponding
density profile for the superheavy nucleus in the spherical ground state is also exhibited
in Fig.~\ref{Fig10}.
It is interesting to note that the density profiles of the two THSIs with $I_z$=81 and $I_z$=208
are nearly the same, as shown indistinguishably in Fig.~\ref{Fig10}.

One observes in Fig.~\ref{Fig10} that the maximum magnitude of the total densities in the
$^{204}120_{184}$ THSIs with $I$=81 and 208 are about the same as those of the nucleus with
a spherical shape. This is in contrast to the case of THSI nuclei in the light mass region
where the maximum density of the THSI nuclei are only about half of the equilibrium nuclear
matter density of the nuclei in the ground state \cite{Sta14}. This arises because the
occurrence of THSI nuclei in the light-mass region is dominated by the nuclear shell effect
and the occupation of the lowest displaced harmonic  oscillator  states with $n_\rho$=$n_z$=0.
For the superheavy nuclei region, the Coulomb repulsion dominates and there are many states
involved. Hence, the nuclear density is not greatly affected by the change from a spherical
shape to a toroidal shape.

The density contours in Figs.~\ref{Fig10} and \ref{Fig11} indicate a well-developed hole
in the density of the nucleus. One can characterize the THSI $^{304}120_{184}(I_z=$81)
by the average geometry parameters of
\begin{equation}
\rho_{\rm max}=0.161 /{\rm fm}^3,~~ R=9.76 {\rm ~fm},~~ d=3.00~{\rm fm},
\end{equation}
which yields $R/d=$3.25. They have the maximum density close to the nuclear matter density,
0.16 fm$^{-3}$. The density profile for the THSI at $I_z$=208 is very similar and will not
be exhibited.

\section{Summary}

Because of the strong Coulomb repulsion, there is a tendency for the shape of a nucleus with
excess charge to bifurcate from a spheroidal into a toroidal shape in the superheavy region.
We examine the case of $^{304}120_{184}$. Without spin, the Coulomb repulsion and
shell effects are not sufficient to allow an equilibrium toroidal shape for $^{304}120_{184}$.
Toroidal minima without spin are possible for superheavy nuclei with greater atomic numbers as
reported earlier \cite{Sta08}.

The spin of a nucleus with an angular momentum about the toroidal symmetry axis has a
stabilizing tendency. We have theoretically located two toroidal high-spin isomeric
states of $^{304}{120}_{184}$ with an angular momentum $I$=$I_z$=81 (proton 2p-2h,
neutron 4p-4h excitation) and $I$=$I_z$=208 (proton 5p-5h, neutron 8p-8h) at $Q_{20}=-297.7$~b
and $Q_{20}=-300.8$~b with energies 79.2 MeV and 101.6 MeV above the spherical ground
state energy, respectively.
The nuclear density distribution of the THSIs $^{304}{120}_{184}$$(I_z$=81 and 208)
have the maximum density close to the nuclear matter density, 0.16 fm$^{-3}$, and
a toroidal major to minor radius aspect ratio $R/d=$3.25 with  $R$=9.76 fm.

Our search to locate the THSIs in $^{304}{120}_{184}$ was focused on the region
-320~b $< Q_{20} <$ -265~b of deformation and it is hard to predict whether two
found toroidal isomers are yrast states. Figure~\ref{Fig9} shows that the
$^{304}{120}_{184}$$(I_z$=81) THSI may appear to lie on the yrast line as there
is no energy minimum with a lower spin lying below this state.
Whether the higher $^{304}{120}_{184}$$(I_z$=208) THSI state is an yrast state is
not known as it depends on the energies of the band of collective states built on
the toroidal intrinsic high-spin state of $^{304}{120}_{184}$$(I_z$=81), by
rotating about an axis perpendicular to the toroidal symmetry axis. A further
investigation is required to study this question.

The results of the single-particle state diagrams and Routhian diagrams obtained
in the present calculations as a function of deformation $Q_{20}$ and the
Lagrange multiplier $\hbar \omega$ indicate that there are shells in the toroidal
shape and the spin degrees of freedom. Extra stability can be maintained at appropriate
occupation numbers, deformations, and spin. Hence, there may be many toroidal
superheavy nuclei as a function of $(Z,N,Q_{20}{\rm ,~and}~ I_z)$ that need to be
uncovered.
The region of toroidal superheavy nuclei may provide an interesting area for further
explorations. Future investigations on ways to produce and to detect these states
with toroidal densities will be of great interest.

%




\begin{acknowledgments}
The research was supported in part by the Division of Nuclear Physics,
U.S. Department of Energy under Contract DE-AC05-00OR22725
and the National Science Center (NCN), Poland, project No. 2016/21/B/ST2/01227.
\end{acknowledgments}



\end{document}